\documentclass[]{aa}  

\usepackage{graphicx}
\usepackage{txfonts}
\usepackage{lipsum}
\usepackage{subcaption}        
\usepackage{lscape}            
\usepackage{placeins}     
                                
\usepackage[breaklinks]{hyperref}
\usepackage{xcolor}
\usepackage{natbib,twoopt}
\usepackage[hyphenbreaks]{breakurl}

\bibpunct{(}{)}{;}{a}{}{,}             
\definecolor{cobalt}{rgb}{0.06, 0.2, 0.65}
\hypersetup{
  colorlinks,
  citecolor=cobalt,
  linkcolor=[rgb]{0.8, 0.2, 1.0},
  urlcolor=cobalt
}

\newcommand{\gs}{\mathrel{\raise0.35ex\hbox{$\scriptstyle >$}\kern-0.5em
\lower0.40ex\hbox{{$\scriptstyle \sim$}}}}
\newcommand{\ls}{\mathrel{\raise0.35ex\hbox{$\scriptstyle <$}\kern-0.5em
\lower0.40ex\hbox{{$\scriptstyle \sim$}}}}

\newcommand{\seco}{\rlap{.} ^s}
\newcommand{\asec}{\rlap{.} {''}}

\begin{document}

   \title{A redshift of $z$\,$=$\,3.20 for the bright arc in eMACS\,J2229.9$-$0808}

   \subtitle{Comment on Wagner \& Falco (2026) ``Hamilton's Object revisited: A challenging source redshift of a strong lensing configuration''}
   
   \author{Ian Smail\inst{1}\corrauth{ian.smail@durham.ac.uk}
     \and Johan Richard\inst{2}\email{johan.richard@univ-lyon1.fr}
     \and H. Ebeling\inst{3}\email{haraldebeling@protonmail.com}
     \and A.C. Edge\inst{1}\email{alastair.edge@durham.ac.uk}
        }

   \institute{Centre for Extragalactic Astronomy, Department of Physics, Durham University, Durham DH1 3LE, UK
     \and Centre de Recherche Astrophysique de Lyon UMR5574, CNRS, Univ Lyon, Univ Lyon 1, Ens de Lyon, F-69230 Saint-Genis-Laval, France
     \and Institute for Astronomy, University of Hawaii, 640 N.\ Aohoku Place, Hilo, HI 96720, USA
}

   \date{Received: May 25, 2026 / Accepted: July 5, 2026}

   \abstract
   {We re-analyse  near-infrared and optical spectroscopy from the Subaru, Gemini, and Keck telescopes of the bright gravitational  arc seen in the $z$\,$=$\,0.62 X-ray cluster eMACS\,J2229.9$-$0808.  The 22 strongest spectral features we identify uniquely determine the redshift of the galaxy as $z$\,$=$\,3.20, as previously reported by \citet{Ebeling25}, not $z$\,$=$\,0.82 as claimed by \citet{Wagner26}.
}


   \maketitle
\nolinenumbers
   
%
%
\begin{figure*}[ht!]
\centering
\includegraphics[width=0.58\linewidth,clip]{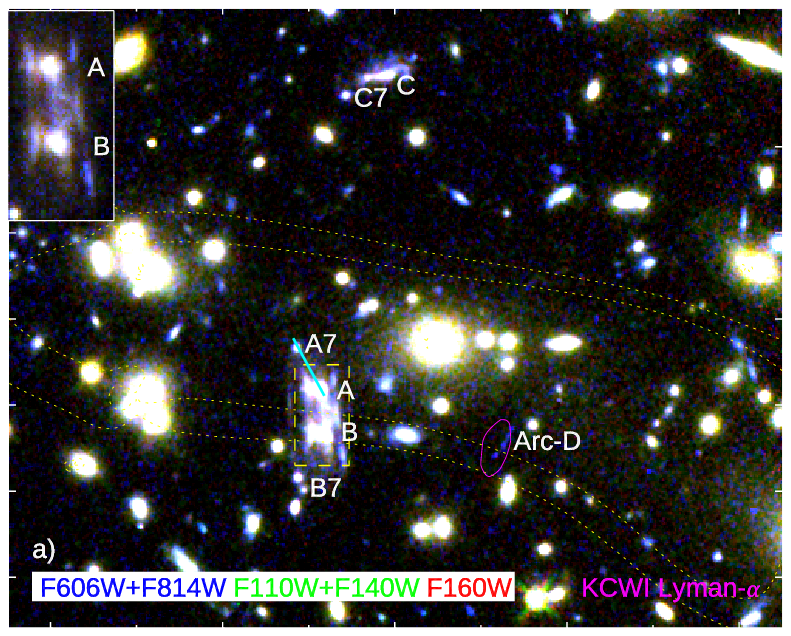} \hspace*{-1.8cm} \vspace*{-1.5cm}\includegraphics[width=0.5\linewidth, bb=15mm 40mm 165mm 120mm]{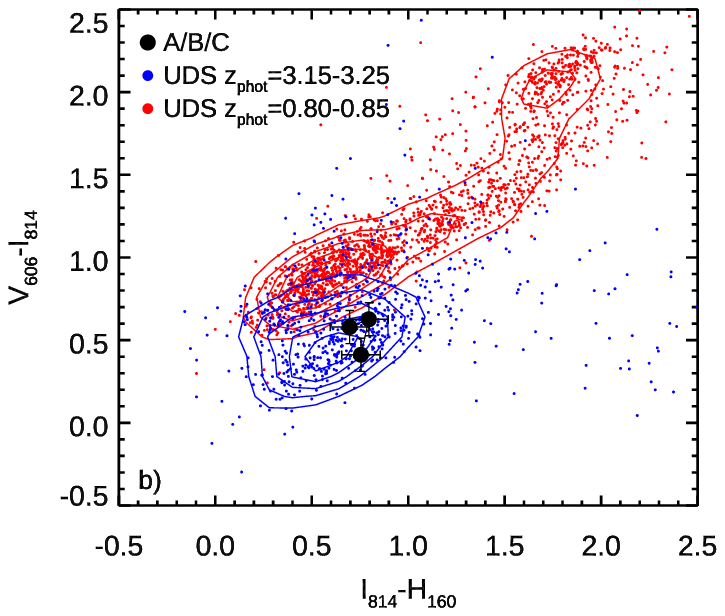} \vspace*{0.3cm}

   \caption{{\bf a)}  A colour image of eMACS\,J2229.9 showing the location of the two images comprising the primary arc (A/B), the counter-image (C), and the faint arc D, identified by \citet{Griffiths21}. The image comprises HST WFC3 and ACS imaging with F160W as Red, F140W+F110W as Green and F814W+F606W as Blue (the major tick marks are 3$''$).   The inset in the top-left panel shows a zoomed image of A/B.  We overlay a (pink) contour showing the bright Lyman-$\alpha$ emission detected with KCWI, claimed to be a LAB by \citet{Griffiths21}, demonstrating that this emission arises from arc D.  We show the aperture used to extract the A+B spectrum from the KCWI IFU cube as a white dashed rectangle and the position of the MOIRCS slit as a cyan line; the GMOS long-slit was orientated to pass through A and B.  The dashed curve shows the $\mu$\,$=$\,50 magnification contour at $z$\,$=$\,3.20 from the \citet{Ebeling25} {\sc lenstool} lens model.
      {\bf b)} $V_{606}-I_{814}$ versus $I_{814}-H_{160}$ distribution of $H\leq$\,26 galaxies  with photometric-redshifts of $z_{\rm phot}$\,$=$\,0.80--0.85 and $z_{\rm phot}$\,$=$\,3.15--3.25 from \citet{Dudzeviciute20}.  We overlay contours of the densities of these two populations and mark the colours of the three images (A, B, and C),  showing that they are consistent with the $z$\,$\sim$\,3.2 population, but 3.2-$\sigma$ discrepant with the colours of $z$\,$\sim$\,0.8 galaxies (cf.\ \citealt{Wagner26}). }
    \medskip
%
%
\centering
\includegraphics[scale=0.55]{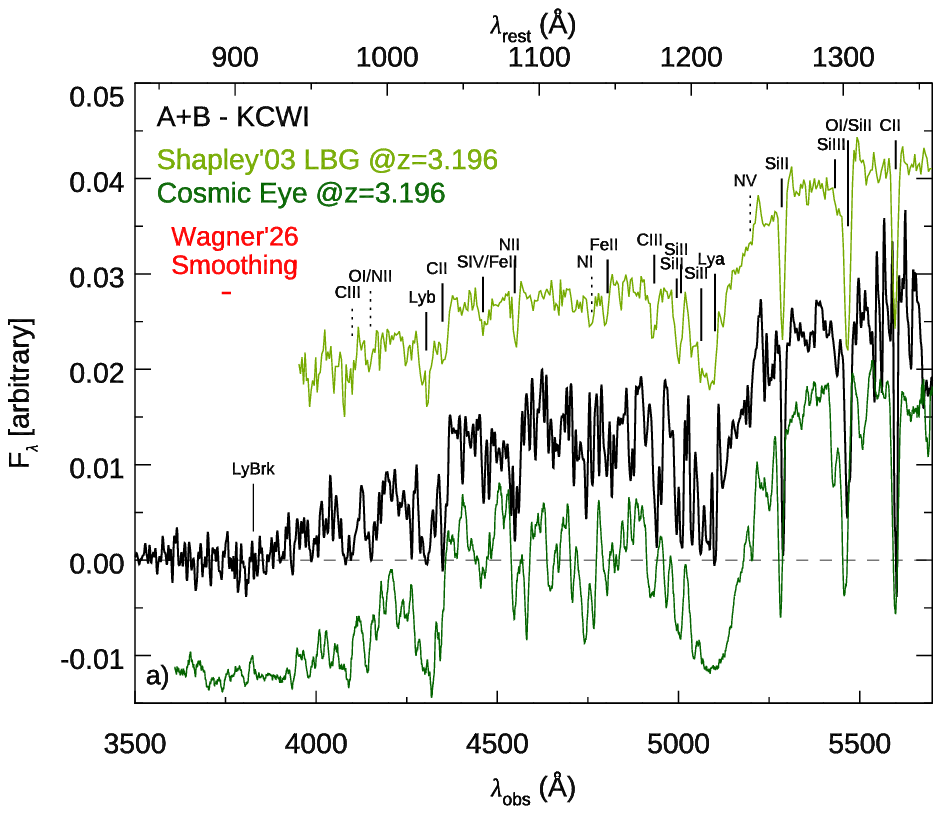}\hspace*{-1.8cm}\includegraphics[scale=0.55]{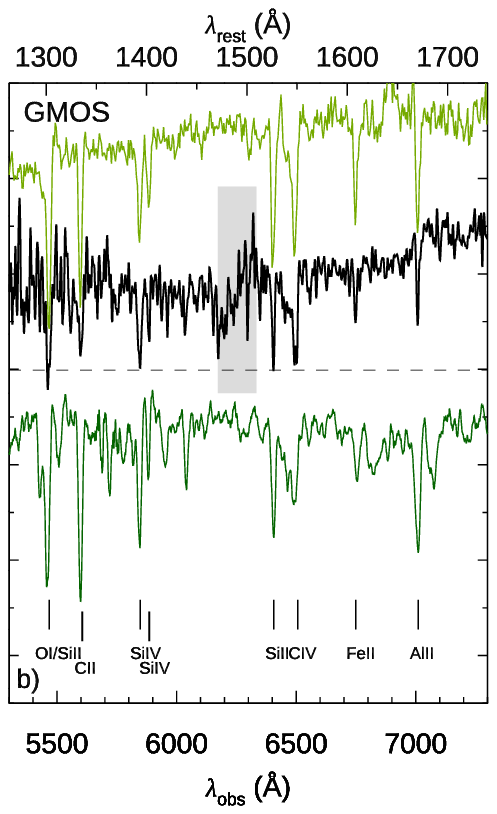}\hspace*{-0.9cm}\includegraphics[scale=0.55]{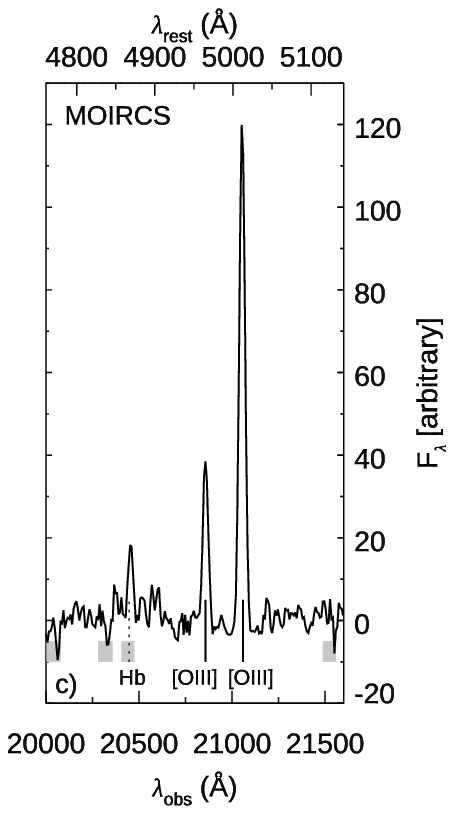}

   \caption{
     {\bf a)}  Our re-reduction of the KCWI spectrum of A+B compared to
the spectrum of  the Cosmic Eye (a bright lensed LBG,  \citealt{Smail07}) and  the composite LBG spectrum constructed by \citet{Shapley03}.   We identify many  absorption features in the spectrum (tentative identifications are marked with dotted lines) and list these in Table~1.  We show the size of the 23.5\AA\ FWHM Gaussian smoothing filter used to subtract the ``continuum'' in the analysis of \citet{Wagner26} as the red line.
      {\bf b)}   Our reduction of the GMOS long-slit spectrum, again compared to the Cosmic Eye and the LBG composite spectrum.  A number of strong absorption features observed in the arc match those seen in the LBG spectra,  most prominently the P-Cygni profile of C{\sc iv}\,1548,1550 at $z$\,$=$\,3.20.   The gray shading indicates regions falling on chip gaps for part of the observations.  
      {\bf c)}  The MOIRCS spectrum of the source discussed in \citet{Ebeling25} showing three emission features corresponding to [O{\sc iii}]\,5007,4959 and H$\beta$ at $z$\,$=$\,3.20.  The gray shading indicate regions affected by  sky emission lines.}
  \end{figure*}

%
%
%
\section{Introduction}

This paper revisits the spectroscopic observations of a bright  lensed  arc seen through the galaxy cluster eMACS\,J2229.9$-$0808 ($z$\,$=$\,0.62, \citealt{Ebeling25}, hereafter eMACS\,J2229.9), which has been variously termed ``Hamilton's Object'' \citep{Griffiths21,Wagner26} or ``The Scream''   \citep{Ebeling25}.

\citet{Griffiths21} analysed GMOS longslit and KCWI integral-field unit (IFU) optical spectroscopy of the three images of the lensed arc and  reported a redshift of $z$\,$=$\,0.82 (which would be noteworthy as this redshift puts the strongly lensed galaxy only slightly beyond the lensing cluster, which they placed at $z$\,$=$\,0.526 based on photometric redshifts).

The redshift for the arc was corrected by \citet{Ebeling25} who used MOIRCS near-infrared spectroscopy along with a re-reduction and re-analysis of the KCWI IFU observations and derived a systemic redshift for the galaxy of $z$\,$=$\,3.201 (they also established the cluster redshift as $z$\,$=$\,0.62 based on the  spectroscopic redshifts of 65 members).

Recently, \citet{Wagner26} have published a re-reduction and re-analysis of the KCWI and MOIRCS observations of the arc, claiming that the former supports $z$\,$=$\,0.82 and the latter is inconclusive.

We present here a comprehensive re-analysis of the available spectroscopy and photometry which demonstrates that they unambiguously determine the  redshift of the lensed galaxy to be  $z$\,$=$\,3.20, as previously reported by \citet{Ebeling25}.

\section{Observations and Data Reduction} 

Imaging of the field of eMACS\,J2229.9 with
Hubble Space Telescope (HST), using  WFC3 (F160W, GO\,13305, PI: C.\ Villforth), ACS (F606W and F814W, GO\,13671, PI: H.\ Ebeling)  and WFC3 (F110W and F140W, GO\,14098, PI: H.\ Ebeling) was obtained in September/October 2013, 2015, and 2016, respectively.  A colour image of eMACS\,J2229.9 constructed from these data (Fig.~1a)  shows  the location of the two images comprising the primary fold arc (A/B, following the naming convention of \citealt{Griffiths21}, at 22$^{h}$30$^{m}$09$\seco$7 $-$08$^\circ$09$'$40$''$, J2000), the counter-image (C), as well as the faint arc (named here D) identified by \citet{Griffiths21}, along with the ``Lyman-$\alpha$ blob'' (LAB) they report (see also \citealt{Wagner26}).  We  measure total magnitudes for A, B, and C from these images and plot their $V_{606}-I_{814}$ and $I_{814}-H_{160}$ colours in Fig.~1b.

The first spectroscopic data obtained of the bright arc was a $K$-band  spectrum of image A  taken on 2016 September 20 using the MOIRCS spectrograph on the Subaru Telescope (PI: H.\ Ebeling).  The source was nodded along the slit for a total exposure of 11.5\,ks; details of the  observational setup and   reduction are provided in \citet{Ebeling25}.  The summed spectrum of A from the nodded pair (Fig.~2c) shows strong [O{\sc iii}]\,4959,5007 emission and potentially H$\beta$ (although this spectral feature lies on a sky line) yielding a robust redshift of $z$\,$=$\,3.201.  
We cannot explain why \citet{Wagner26} were unable to extract a similar spectrum from their attempt to re-reduce the MOIRCS observations.

On 2018 August 14, a longslit spectrum of A/B was obtained  with the Gemini-N GMOS spectrograph (program GN-2017A-Q-48, PI: R.\ Griffiths).  The data we analyse totalled 5.4-ks integration using the R400 grating with a central wavelength of 7000--7100\AA.  Data and associated calibration frames were retrieved from the Gemini archive and reduced using the DRAGONS pipeline \citep{Labrie23}, including standard bias subtraction, flatfielding, image combination, spectral extraction, wavelength calibration and fluxing.   The resulting spectrum (Fig.~2b) shows several strong absorption lines, including a broad absorption feature at $\sim$\,6500\AA, which we identify as the C{\sc iv}1548,1550 doublet at $z$\,$\sim$\,3.20.

The GMOS data were not re-analysed by \citet{Wagner26}, who just listed the
features that \citet{Griffiths21} stated were detected in this spectrum:   [O{\sc ii}]\,3727,3729,  Ca H\&K and CN\,4180\AA\ absorption.
However, we see no compelling evidence in the spectrum (Fig.~2b) for an [O{\sc ii}] emission line at $\sim$\,6783\AA, or Ca H\&K absorption lines at 7158 and 7224\AA,  corresponding to $z$\,$=$\,0.82.   We also suggest that the claimed CN\,4180\AA\ absorption, which would lie at $\sim$\,7610\AA, likely corresponds to a residual due to the A-band telluric absorption.

Finally, an inconclusive Gemini-N GMOS R150 IFU spectrum was obtained in 2017 May and June (see \citealt{Griffiths21}), but was superseded by  a 15.6-ks Keck-II KCWI IFU observation of the cluster core obtained on  2020 September 16 (H315, PI: R.\ Griffiths), which covered A/B/C and the faint arc D.  Full details of the observations are given in \citet{Griffiths21}; this spectrum is also re-analysed and discussed in \citet{Wagner26}.  We retrieved the raw data cubes and associated calibrations from the Keck archive and these were re-reduced  using the PypeIt reduction pipeline \citep{Prochaska20} to obtain a calibrated data cube.  We then extracted spectra corresponding to A+B, C, and the  arc D from the cube. We show the spectrum of the fold arc A+B (which has the highest signal-to-noise) in Fig.~2a and overlay in Fig.~1a a contour showing the extent of the bright emission detected at 5106.9\AA\ in the IFU data cube in the vicinity of arc D.

\section{Analysis and Results} 

\citet{Ebeling25}  briefly reported the results of our re-analysis of the KCWI observations and listed the identification in the  spectrum of A+B of some of the stronger features, including Lyman-$\alpha$; Lyman-$\beta$; the Lyman break; Si{\sc ii}\,1260; O{\sc i}\,1302; and C{\sc ii}\,1335 interstellar medium (ISM) lines at $z$\,$\sim$\,3.20, in agreement with the redshift derived from the [O{\sc iii}] lines in their MOIRCS spectrum.  They noted that these spectral features  had been  previously identified by \citet{Griffiths21} as  Mg{\sc ii}\,2796,2803; Fe{\sc ii}\,2365;  not noted;  Fe{\sc ii}\,2905; Fe{\sc i}\,3004; and Ti{\sc ii}\,3074, respectively.

Recently, \citet{Wagner26} have also re-reduced the KCWI observations using {\sc PypeIt} and re-analysed them.  However, their analysis of the spectrum of the lensed galaxy involved subtracting a smoothed version  (using a  Gaussian filter with a FWHM of 23.5\,\AA) from the data,  before attempting to identify features in the residuals.  They tabulate 17 absorption features, including several reported by \citet{Griffiths21}, as well  as 10 emission features, identifying many of these with low-equivalent width  lines that are typically only visible in extremely deep integrations or stacked spectra, 
and state that these support a redshift of $z$\,$=$\,0.82 for the source.

We do not give a detailed analysis of the features reported by \citet{Wagner26}, nine of which appear to correspond to real absorption lines in Table~1 (although four of these lack a proposed identification at $z$\,$=$\,0.82 in \citealt{Wagner26}).  Instead,
we note that the scale of the 23.5-\AA\ FWHM  filter used by them to create a ``continuum-subtracted'' spectrum,  is far narrower than some of the real features in the spectrum (Fig.~2a).   We also repeat their processing steps on the three spectra plotted in Fig.~2a and show these in Fig.~3 (which can be compared to Fig.~4 in \citealt{Wagner26}).
It is evident  that the subtraction of such a smoothed version of the spectrum from itself  not only removes real information from broad features, but also  significantly confuses the subsequent identification and interpretation of real features.     We therefore focus our analysis on the features visible in the spectra shown in Fig.~2.

%
%
 \begin{figure}[ht!]
\centering
\includegraphics[width=\linewidth,clip]{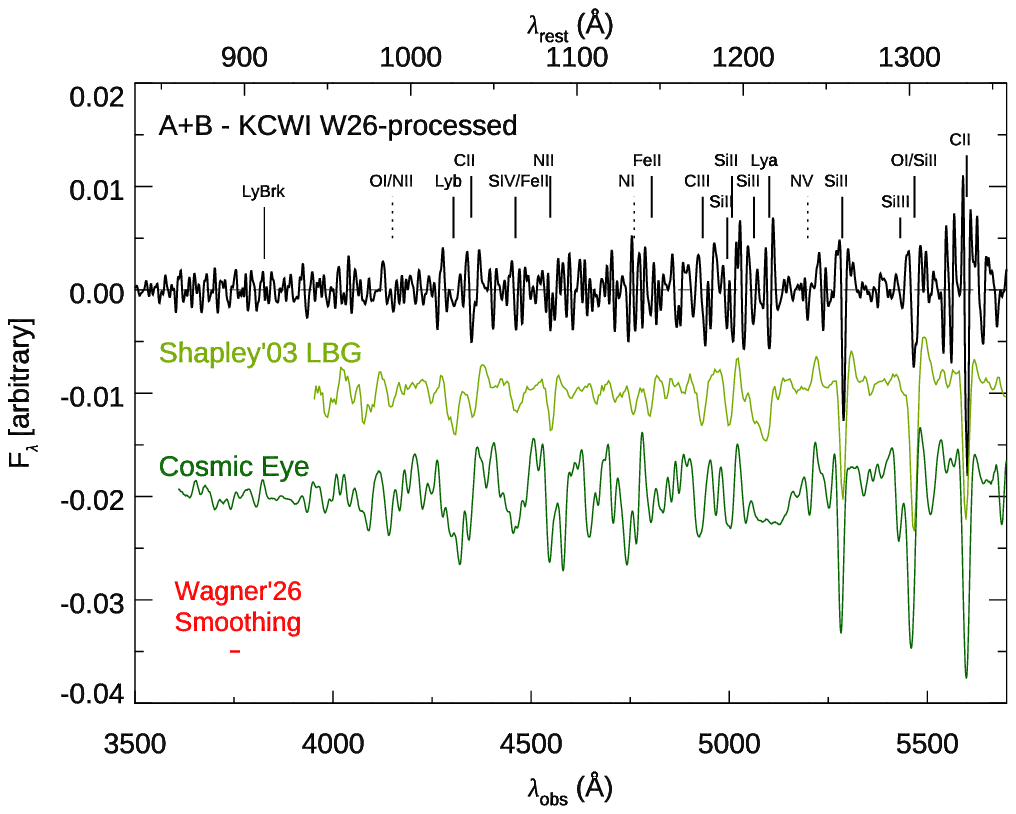} 
   \caption{The three spectra from Fig.~2a after applying the processing used in \citet{Wagner26}.  This involves subtracting a 23.5-\AA\ FWHM smoothed version of the spectrum from itself (the filter size is shown in red) and then convolving the residuals with a 4.7-\AA\ FWHM Gaussian.  We mark the features that were visible in the original spectra in Fig.~2a. Only the narrower, strongest and most isolated of these survive this process.}
 \end{figure}

The combined optical spectroscopy of the arc from KCWI and GMOS (Figs.~2a \& 2b) shows a large number of  absorption features. These correspond to  strong restframe UV absorption lines, most of them  common ISM lines seen in Lyman-Break Galaxies (LBGs) \citep[e.g.,][]{Shapley03,Steidel16}.   To demonstrate this, we  compare our spectrum to that of another bright lensed LBG, the Cosmic Eye from \citet{Smail07}, and to the composite LBG spectrum from \citet{Shapley03}.   The similarity of these three spectra is striking:    many of the absorption lines are clearly seen in all three spectra, including Lyman-$\alpha$ and Lyman-$\beta$, as well as a host of metal ISM lines.     In total we identify 28  spectral features and mark these in Figs.~2, listing the stronger lines in Table~1.

We determine a  systemic redshift of $z$\,$=$\,3.2013\,$\pm$\,0.0002 for the lensed galaxy from the restframe optical emission lines, compared to the median redshift of $z$\,$=$\,3.1960\,$\pm$\,0.0005 derived from the restframe UV absorption lines (Table~1), i.e., the UV lines are offset by $-$380\,$\pm$\,40\,km\,s$^{-1}$. Such blue shifts are frequently seen in the ISM absorption lines in LBGs and arise from outflows \citep[e.g.,][]{Shapley03}.

We also confirm that the broadband colours of the lensed galaxy are  consistent with a redshift of $z$\,$=$\,3.20, rather than   $z$\,$=$\,0.82 (\citealt{Griffiths21}; \citealt{Wagner26}), using the $V_{606}-I_{814}$ versus $I_{814}-H_{160}$ colour-colour diagram in Fig.~1b.  A comparison of the colours of $H$\,$\leq$\,26.0 field galaxies with photometric redshifts of $z_{\rm phot}$\,$=$\,0.80--0.85 and $z_{\rm phot}=$\,3.15--3.25 from the {\sc magphys} analysis of the UKIDSS UDS DR11 \citep{Lawrence07} catalogue\footnote{For a full description see:\\ \url{https://www.nottingham.ac.uk/astronomy/UDS/DR11/}}  by  \citet{Dudzeviciute20}, shows the two populations to be fairly distinct.  We mark the colours of the three arcs (A, B and C),  to illustrate that they are consistent with the $z$\,$\sim$\,3.2 population and 3.2-$\sigma$ discrepant with the colours of $z$\,$\sim$\,0.8 galaxies based on a likelihood-ratio test.

\section{Discussion}

The 22 strongest  absorption and emission lines listed in Table~1 and plotted in Fig.~2 show conclusively that the bright lensed galaxy in eMACS\,J2229.9 has a redshift of $z$\,$=$\,3.20, as first reported by \citet{Ebeling25}, not $z$\,$=$\,0.82 as claimed by \citet{Wagner26}.

In Fig.~1a we plot the contours corresponding to a magnification of $\mu$\,$=$\,50 as dashed lines, showing that the fold image of the arc (A/B) is highly magnified.  From the {\sc lenstool} lens model presented in \citet{Ebeling25}, we estimate that A+B  have a total area-weighted magnification of 57\,$\times$ (indicating an intrinsic brightness of $H_{160}$\,$\sim$\,24.6).  Searching the UDS galaxy catalogue used in Fig.~1 for $z$\,$\sim$\,3.2 galaxies with $H_{160}$\,$\sim$\,24.6 and  colours matching the arc, we  use their physical properties, derived from  {\sc magphys} modelling of their $\sim$\,20-band photometry by \citet{Dudzeviciute20}, to estimate the properties of the galaxy.  This suggests that  the galaxy is likely to be a strongly star-forming, but sub-M$^\ast$, system with a star-formation rate of approximately, $\sim$\,15\,$\pm$\,5\,M$_\odot$\,yr$^{-1}$ and a stellar mass of $M_\ast \sim (4\pm2)\times10^{9}$\,M$_\odot$.

We  overlay on Fig.~1a a contour showing the bright Lyman-$\alpha$ emission detected in the KCWI observations at 5106.9\AA\ (corresponding to $z$\,$=$\,3.201) which is reported as a  LAB by   \citet{Griffiths21}, see also  \citet{Wagner26}.  This emission is coincident with the highest-magnification part of the faint arc  D,  a connection which had not previously been made,  indicating that it too lies at $z$\,$=$\,3.20.  From the $z$\,$=$\,3.2 magnification contour we see that this source (which is only visible in F606W and F814W) is  highly magnified,  with a magnification of $\sim$\,50--60\,$\times$, corresponding to an intrinsic brightness of $V_{606}$\,$\sim$\,30 and suggesting it is probably a  low-mass galaxy.   However, the observed Lyman-$\alpha$ size of this source likely does not reflect its intrinsic extent, but results instead from the significant shear of the image and the low spatial resolution of the wide-field KCWI observations.  Whether this source is an intrinsically  spatially extended LAB thus remains unclear.  

We  note that the identification of arc D as the claimed ``LAB''  refutes one argument made by \citet{Wagner26} against the bright arc lying at  $z$\,$=$\,3.20.  Their claim  was that this would require that the $z$\,$=$\,3.20 ``LAB'' should show a similar multiple image configuration to that displayed by bright arc images, A/B/C, but  that only one image of the LAB was seen within the KCWI field.   However, our identification of the bright Lyman-$\alpha$ emitter as D removes this objection, as D actually comprises two highly magnified images in a fold configuration.   The predicted counter image is expected to be $\sim$\,10\,$\times$ fainter, making it undetectable in the existing HST imaging (observed $V_{606}$\,$\sim$\,28), and is located near 22$^h$30$^m$08$\seco$72 $-$08$^\circ$09$'$22$\asec$3, which falls outside of the  KCWI footprint.

\section{Conclusions}

We conclude that the redshifts  for  the lensed arc (A/B/C, $z$\,$=$\,0.82) and the cluster (a photometric redshift of $z$\,$=$\,0.526), originally proposed in \citet{Griffiths21}, and  reiterated in \citet{Wagner26} are both in error. \citet{Ebeling25} has provided extensive spectroscopy of cluster members showing that the cluster redshift is $z$\,$=$\,0.62.   They also reported a redshift for the  bright fold arc  of $z$\,$=$\,3.20,  correcting the value reported in \citet{Griffiths21},  
a fact that has been  disputed by \citet{Wagner26}. Our re-reduction and re-analysis of all of the spectroscopy of this source uncovers 22 robust spectral features which show definitively that the lensed galaxy has a redshift of $z$\,$=$\,3.20.  This galaxy thus has the same redshift as the highly magnified arc D that we identify as a low-mass Ly$\alpha$ emitter and which  lies no more than a few 10's kpc in projection from the brighter galaxy in the source plane.   

Finally, we note that the issue of redshift ambiguity discussed by \citet{Wagner26} is a potential problem \citep[e.g.,][]{Smail05}, especially for galaxies at $z$\,$\sim$\,2 when relying solely on visible/red spectroscopy. However, there is no ambiguity in the redshift of the eMACS\,J2229.9 arc when the spectra are appropriately analysed: it lies at $z$\,$=$\,3.20.

\begin{acknowledgements}
We thank the Referee for their rapid, helpful report.
We also thank Ryan Cooke for reducing the KCWI spectrum.  
\end{acknowledgements}

\bibliographystyle{aa_url} 
\bibliography{TheScream}

%
%
\begin{table}
  \centering
  \caption{Robust  line identifications}
  \begin{tabular}{cccl}
    \hline
    $\lambda_{\rm obs}$ & $\lambda^{\rm ID}_{\rm rest}$ & $z$ & ID \\
    \hline
    \hline
4303.7 & 1025.7 & 3.19588 & Ly$\beta$ \\
4350.1 & 1036.3 & 3.19775 & C{\sc ii} \\ 
4461.5 & 1062.7 & 3.19825 & S{\sc iv}/Fe{\sc ii} \\  
4548.9 & 1084.0 & 3.19636 & N{\sc ii} \\
4801.4 & 1145.0 & 3.19337 & Fe{\sc ii} \\  
4938.5 & 1176.0 & 3.19941 & C{\sc iii} \\   
4993.9 & 1190.4 & 3.19514 & Si{\sc ii} \\ 
5008.5 & 1193.3 & 3.19718 & Si{\sc ii} \\  
5061.5 & 1206.5 & 3.19515 & Si{\sc iii} \\ 
5101.3 & 1215.7 & 3.19616 & Ly$\alpha$ \\  
5288.7 & 1260.4 & 3.19605 & Si{\sc ii}  \\   
5436.4 & 1296.5 & 3.19316 & Si{\sc iii}/C{\sc iii} \\ 
5467.5 & 1303.7 & 3.19385 & O{\sc i}/Si{\sc ii} \\   
5600.0 & 1334.5 & 3.19630 & C{\sc ii}  \\ 
5845.9  & 1393.8 & 3.19418 & Si{\sc iv} \\
5886.7 & 1402.8 & 3.19642 & Si{\sc iv} \\ 
6404.6 & 1526.7 & 3.19507 & Si{\sc ii} \\ 
6497.7  & 1549.0 & 3.19478 & C{\sc iv} \\
6747.6  & 1608.5 & 3.19494 & Fe{\sc ii} \\ 
7007.6   & 1670.8 & 3.19417 & Al{\sc ii} \\ 
20835.7 & 4959.0 & 3.2016 & [O{\sc iii}]  \\
21034.6 & 5007.0 & 3.2010 & [O{\sc iii}]  \\
    \hline
\end{tabular}
\end{table}

\end{document}